\documentclass[acmsmall]{acmart}

\AtBeginDocument{%
  \providecommand\BibTeX{{%
    \normalfont B\kern-0.5em{\scshape i\kern-0.25em b}\kern-0.8em\TeX}}}

\usepackage{wrapfig}
\begin{document}

\title[Work Online, Welfare Calls, and Wine Night]{Work Online, Welfare Calls, and Wine Night: Effects of the COVID-19 Pandemic on Individuals' Technology Use}

\author{Bill Tomlinson}
\email{wmt@uci.edu}
\affiliation{%
  \institution{University of California, Irvine and Victoria University of Wellington}
  \streetaddress{5068 Bren Hall}
  \city{Irvine}
  \state{CA}
  \postcode{92697}
}
\author{Rebecca W. Black}
\email{rwblack@uci.edu}
\affiliation{%
  \institution{University of California, Irvine}
  \streetaddress{5072 Bren Hall}
  \city{Irvine}
  \state{CA}
  \postcode{92697}
}

\renewcommand{\shortauthors}{Tomlinson and Black}

\begin{abstract}
The COVID-19 pandemic has changed the ways many people use computational systems.  We conducted an empirical study, using qualitative and quantitative analyses of free-response surveys completed by 62 US residents, to explore how COVID-19 affected their computer use across work, education, home life, and social life.  Nearly all participants experienced an increase in computer usage for themselves or a family member in one or more of the four domains. The increases involved both increasing frequency of existing uses as well as the adoption of new types of use. Changes in usage impacted many aspects of people's lives, including relationships, affective experiences, and life trajectories. Understanding these changes is important to the future of HCI, as the field adapts to COVID-19 and potential future pandemics.

\end{abstract}




\maketitle

\section{Introduction}
At the time of writing (mid September 2020), COVID-19 (hereafter, COVID) has caused the deaths of over 900,000 people globally, with several thousand more people dying each day \cite{ourworld}. The virus has led to profound suffering for those killed and their families, as well as for those who survived and yet have enduring effects such as lung scarring and heart damage \cite{berkeley}.

Govermental, corporate, and community responses to this pandemic have led to dramatic changes in how people in many countries live their lives.  Governments have enacted lockdowns and quarantines \cite{WHO}, businesses have adopted work-from-home policies \cite{Fung}, and schools have shifed education online \cite{schools}.  Beyond the physical effects of the COVID illness, these and other types of societal responses have also led to major impacts on people.

Computational systems are deeply embedded in industrial civilizations, with billions of people using mobile phones \cite{phones}, social networking systems \cite{social} and other platforms.  The changes in individuals' daily lives due to COVID also resulted in major shifts in how these individuals use computational systems.  While these changes extend to all parts of industrial societies, from the operation of health infrastructure \cite{weforum} to modeling environmental impacts \cite{davis}, in this paper we explore in particular how individuals' computer use has been altered as a result of COVID.

Specifically, we seek to explore the following research questions:
\begin{itemize}
\item RQ1: In what ways has the COVID-19 epidemic shaped computer use in an industrial country, across the domains of work, education, home life, and social life?
\item RQ2: What activities, devices, and platforms were most salient in people's usage across these domains?
\end{itemize}

We take an exploratory, human-centered approach to this work, focusing primarily on people's lives, rather than the technologies involved.  Hence we have framed the research around work, education, home, and social life, four critical domains in the lives of many individuals living in industrial civilizations, rather than framing the work around particular technological platforms such as videoconferencing, social media, and mobile phones. We chose these particular dimensions due to the lived experiences of the authors over the past several months, who identified these issues as most salient in their own experience of COVID; nevertheless, based on the findings from this study, additional domains emerged that would be useful to explore in future work.

We conducted an empirical investigation of these domains, via a survey on Amazon's Mechanical Turk platform (hereafter AMT, due to this platform's problematic naming, unpacked by Aytes \cite{aytes}).

Findings ranged from increases in computer use for work-related videoconferencing, online educational activities, gaming, and social media, to more surprising patterns such as the ways people used technology to rekindle relationships and pursue new opportunities in the COVID era.

This paper makes an empirical contribution to the human-computer interaction (HCI) body of knowledge by providing timely findings regarding a current, global disruption, based on data collected as this disruption was unfolding.  As such, it provides a baseline against which future studies of the effects of COVID on computer usage may be compared.

COVID is changing how people in the industrialized world use computers in their everyday lives.  This paper provides concrete evidence for how this change took shape in the first few months of the pandemic.

\section{Related Work}
Many scholars have studied the ways in which computing has factored into the health-care implications of COVID, e.g.,  \cite{JAVAID2020419}. This topic has also been explored in the popular press \cite{Koeze}, and by governmental bodies such as the CDC \cite{CDC}.

Various scholars have begun to explore the role of HCI in the context of the COVID pandemic. Perhaps most broadly, Yvonne Rogers \cite{rogers} identified many of the abundant ways in which COVID is changing how people engage with computers.  Her reflective piece provided rich descriptions of her own experiences, and the experiences and actions of others around the world. Similarly, Mikael Wiberg \cite{wiberg} identified work, education, and family as three domains of social interaction that were moving online as a result of COVID. In this paper, we seek to build on their recognition of the abundant ways that COVID is changing people's lives, and to provide deeper insight into these myriad phenomena.   

In his column in Communications of the ACM \cite{cerf}, Vint Cerf also contemplated the contributions that computing could make to ``post-COVID society''.  We very much agree, and here explore the ways that people in an industrial civilization have been adopting and adapting existing technologies to fill new needs and wants.

Going deeper into the specific areas that lie at the juncture of HCI and COVID that we sought to explore with this paper, we here identify a range of work relevant to this paper's contributions.

\subsection{Computer-Supported Work and COVID}
Scholars in computer-supported work have explored numerous topics relating to working from home; past research efforts hav explored remote team members' experiences \cite{Koehne}, the social practices of working from home \cite{Venolia}, and appropopriate management styles to support those who work from home due to disability \cite{spark}. While these and many other investigations have helped enable working from home over the past years, research in computer-supported work has only just begun to engage with issues related to COVID; we identify the most salient research projects here.

Ren et al. \cite{ren} held a workshop at the DIS '20 conference to design for healthy behaviors given the rise in working from home as a result of COVID. Lottridge \cite{lottridge} identified how video chatting, including in the context of working from home but in other aspects of their lives as well, is a key facet of the COVID experience for many people. Ahmed et al. \cite{ahmed} explored the implications of COVID for one particular workforce---garment workers in Bangladesh. And Steed et al. \cite{steed} discussed challenges with conducting AR/VR research when in-person user studies and the sharing of devices are no longer viable. While all of these papers investigate particular facets of work experiences during COVID, none provided a context in which people could self-identify the factors most important to them to discuss in this context.  This paper seeks to fill that gap.

\subsection{Computer-Supported Learning and COVID}
Computer-supported learning has a similarly long history to that of computer-supported work, including work to build a sense of community among distance learners \cite{sun, Ngoon}, shared context in online learners \cite{Hamilton}, students' experiences and future directions in massive open online courses \cite{Zheng, Scanlon}, and cross-cultural factors in online learning \cite{Morales}.  

In this domain as well, scholars have begun to grapple with COVID. For example, Grudin \cite{grudin} described the challenges education faced when ``[f]orced by Covid-19 into functioning remotely'', and reflected on the potentially applicability of research from the ``distant past'' (20 years previous). Pal et al. \cite{Pal} investigated the impact of multimedia quality on learners' experiences in online education resulting from COVID. Triyason et al. \cite{triyason} identified difficulties that arose in efforts to created hybrid classrooms in a university setting in response to COVID.  And Xu and Chen \cite{xu} investigated factors that contributed to the sense of security of university students in response to the COVID outbreak.  Similar to the distinction between the work presented here and previous research in computer-supported work, we see the contribution of this paper  involving an investigation of what people themselves found to be the most salient aspects of their educational experiences in the time of COVID.

\subsection{Computer-Supported Home Life and COVID}
People use computers for many different aspects of their home life. Scholars have investigated a range of aspects of home computer use, including the use of wearables \cite{Huang} and games \cite{Gerling} in home health care contexts, how children search at home \cite{Druin}, and how people find informal technical support at home \cite{Poole}. Some of these overlap with topics explored in work and learning above (e.g., the use interactive multimedia systems). 

Recently, scholars have begun to study various aspects of how COVID affects the role of computing at home, including how technology can be used to support the well-being of children \cite{goldschmidt_covid-19_2020} and the impact of COVID on elders' computer use \cite{Morrow}. Studies have been done on the role of gaming during COVID \cite{amin_online_2020}, and in particular on problematic online gaming \cite{king}. Problematic internet use during COVID has also been studied \cite{KIRALY2020152180}. Nevertheless, here, as with work and education, we could find no studies that allowed people to identify the changes most salient to them.

\subsection{Social Computing and COVID}
Social computing is also a major domain of ongoing research, with scholars exploring how current and potential future computing systems can help families separated by distance to remain connected \cite{Ataguba, Chai}, predicting social interactions in online games \cite{Frommel}, and investigating how social media users engaged in social comparison \cite{Burke}, among many other topics.

Researchers in this domain are beginning to focus on COVID-related work as well. For example, Mejova and Kalimeri \cite{Mejova} investigated representations of COVID in advertisements on Facebook. Priego and Wilkins \cite{Priego} used the creation of comics to explored people's experiences engaging with elderly relatives via videoconferencing. Qazi et al. \cite{Qazi} analyzed 524 million tweets relating to COVID that were posted in the 90 days following February 1, 2020. Iqbal and Campbell \cite{iqbal} looked at the need created by COVID for touchless interactions.  And Laato et al. \cite{LAATO2020101458} investigated whether location-based games such as Pokemon Go caused players to socialize during the COVID pandemic. Again, none of these research efforts sought to elicit the participants' own opinions about how their computer usage had changed in this domain.

Since we were unable to find any studies that provided people with the opportunity to discuss the topics most relevant to them across any of the four topics described above, we believe that this study provides a unique contribution to the scholarly literature.

\section{Methodology}

\subsection{Survey}
Given the lack of previous research providing an opportunity for residents in an industrial nation to describe the computing-related topics most relevant to them as a result of COVID, we set out to study this topic. We created an online survey instrument in Qualtrics. This survey included several demographic questions and four questions about computer use during the pandemic.

Demographic questions included the following: household income (using a MacArthur Foundation protocol \cite{macarthur} updated with income ranges drawn from the US Census \cite{census}), political views (using a Pew Research Center protocol \cite{pew}), education level (using a SAGE Encyclopedia of Communication Research Methods protocol \cite{Allen}), and gender (using best practices in the field of HCI \cite{Spiel:2019:BGS:3342541.3338283}, allowing participants to select any or all of the following options: ``Woman'', ``Man'', ``Non-binary'', ``Prefer not to disclose'', or ``Prefer to self-describe'' (with the last option providing a free-response field)).

The four questions about computer use during the pandemic were structured similarly: 
\begin{itemize}
\item  Has the COVID-19 (aka coronavirus) pandemic changed how you or a family member has used a computing system or digital media for their <work/education/home life/social life>? Please, in at least 100 characters, describe any changes or lack of changes.
\end{itemize}

This study's methodology was approved by the authors' university's Institutional Review Board (IRB).

\subsection{AMT Study}
We conducted a survey of US residents via AMT.  AMT is an online system by which individuals and organizations can post ``Human Intelligence Tasks'' (HITs), and workers can complete those tasks in exchange for payment.

\subsection{Fair Payment/Treatment}

Participants were paid in line with best practices in crowdsourcing research in HCI \cite{Silberman}. Based on the median response time for participants in a pilot study, we established that the task took less than five minutes for most participants; given a target pay rate of US\$15/hour, the payment for the task was set at US\$1.25 per participant. However, after running the full survey, we found that the median time to completion was 9.2 minutes; therefore, AMT's bonus mechanism was used to increase the average rate of pay to US\$15/hour.

\subsection{Runs Conducted}
We conducted a series of pilot studies to refine the experimental procedure, including revisions to the AMT process, the Qualtrics survey, and the interface between the two.

We ran the final survey on May 30, 2020.  The cultural context of this time period is of substantial relevance to this study.  The people of the world, and in particular the US, had been coping with the COVID pandemic for several months.  In addition, protests against the police murder of George Floyd on May 25, 2020 \cite{Johnson}, were just beginning.  This was a period of significant anxiety and distress for many people in the US.

We initially recruited 100 participants.  However, one participant emailed us that she had accidentally hit the submit button before completing the survey, and asked if we could allow her to retake the survey, which we did.  Therefore, we ended up with 101 completed surveys.  

\subsection{Excluded Data}
Of the 101 responses we collected, we excluded 39 participants who did not complete the survey in good faith according to our quality control metrics, e.g., who pasted text from the internet rather than writing original content.   Results reported here are based on the remaining 62 completed surveys.
	
\subsection{Qualitative Coding}

Once all participants completed the study, we analyzed the demographic and qualitative data.  In the first stage of coding, we engaged in an iterative process of applying descriptive codes to the data, meeting to discuss and refine codes, and then recoding \cite{saldana}. In the next cycle of coding, we looked for patterns in the data and generated thematic categories that represented areas in which the codes converged or diverged in notable ways. 

To facilitate pattern recognition, we also used ``quasi-statistical'' methods \cite{Becker}. This involved making a count of all codes and using these numbers ``to facilitate pattern recognition or otherwise to extract meaning from qualitative data, account for all data, document analytic moves, and verify interpretations.'' \cite{Sandelowski} 

\section{Results}

\subsection{Demographics}
Participants included in the analysis lived in 20 different US states.  Their average age was 40.2 years old. Forty identified as men, 22 as women, and one as both a woman and a man.  Thirty-three participants identified as liberal or very liberal, 13 as moderate, and 17 as conservative or very conservative. The median reported household income was \$35,000 through \$49,999, with several reporting income less than \$15,000, and one reporting income over \$200,000.  The majority held bachelor's degrees, with a range of other educational levels represented as well, from high school diplomas to advanced degrees.

\subsection{Changes in Usage}
Here we describe findings that relate to RQ1, about changes that participants experienced in computer usage regarding work, education, home life, and social life. Overall, we found that the vast majority of participants reported an increase in computer usage for themselves or a family member in at least one of the four domains.  Some reported increases in all four domains, while others in only a subset of them.  

Only one participant\footnote{Responses from two additional participants were ambiguously worded and not included in this portion of the analysis.} did not report an increase in their computer usage in any domain. This participant, a 34 year old man from Nevada, said no aspect of his life had changed and called into question the legitimacy of COVID. ``Our family is living normal and refuse to fall for this fake ass trash virus''.\footnote{All quotes are presented exactly as written by study participants, including grammatical errors, irregular punctuation, etc.} 

Only one participant claimed to have experienced any form of decreased computer use as a result of COVID, and it was a result of an increase in her computer usage for work.  This participant, a 42 year old woman from Alabama, wrote: ``I have had to turn my personal computing system at home into my tiny office.  I no longer have the time or inclination to casually browse the internet or any other such thing at home now.  As soon as I can get away from the computer, I stay away as much as possible.''


Below, we address separately the qualitative findings in each of the four domains: work, education, home life, and social life.

\subsubsection{Work}
A large number of participants identified working from home as a key change in their experiences as a result of COVID.

Some people used their own devices to work from home, sometimes augmenting their existing systems. A 31 year old man from New York wrote: ``I've had to work from home, and so have set up a remote workstation in my living room. I've added an extra monitor to my desk and an external keyboard and mouse, whereas before I would have just used my laptop on its own when working from home.'' 

Other people had devices provided for them by their work.  A 56 year old woman from Ohio wrote: ``My husband now works from home. He received a laptop from work so our internet time has gone up.''

Many people who were working from home found that their computer usage increased dramatically. For example, a 36 year old woman from California wrote: ``I use the computer a lot more since I work at home now. I use my computer pretty much all day long.''

Nevertheless, some also found that working online compromised their ability to work effectively. For example, a 27 year old person from New York who identified as both a woman and a man wrote: ``I have been using the computer to telecommute to work. I have been calling and video chatting patients and clients to perform evaluations and consultations with them. This has been less effective because I think that an initial in-person visit for evaluation would be better as you cannot do physical examinations over the internet.''

A substantial number of participants noted that nothing or very little had changed about their work lives, despite COVID. For example, a 44 year old woman from Virginia wrote: ``I am required to go to my work site every day. We had one zoom meeting while at work the other day, but besides that there has been no change in the way I do my work.''

For those who experienced no change, technology still often played a role.  For example, this 26 year old man from Maryland wrote:
``[T]he pandemic has not changed how anyone in my household uses computers for their work. I worked full time on MTurk prior to the pandemic, and I'm still doing the same during the pandemic, so that has not changed. No one else in my household has a computer, and I help my father with his work by occasionally sending emails or printing files on my computer, and the pandemic has not changed that.''

There was ambivalence within the study population about whether the increase in computer use for work as a result of COVID was a good thing or bad. A 30 year old man from New York wrote: ``I use [computers] 18 hours a day now and work completely from home. It is a great change that I hope is permanent.''

Similarly, a 30 year old man from Washington state commented that his wife was enjoying the change: ``The happenings of COVID-19 have made it necessary for my wife to extremely use the computing technology for her various work tasks.She used to work from office but this has definitely changed as a result of the lock down.She currently works from her computer here at home and she kind of likes the whole idea.I would say that her use of the computer has greatly increased compared to sometimes back.She is able to meet clients online and efficiently close business where necessary.''

A 62 year old woman from Illinois noted that technology was (mostly) helpful: ``from this COVID-19 issue, family members are widely used computing system or digital media for their work. it is mostly useful and helpful.''

Conversely, a 26 year old man from Florida wrote: ``I have had to change to working from home. I have to wait until September to go back to work. its annoying and a big inconvenience to myself.''

Across these disparate experiences that participants reported, we saw patterns of both increased and sometimes novel technology use. Many participants expressed a sense of necessity related to their work computer use. Nevertheless, some participants saw the changes in their work usage as a positive outcome.

\subsubsection{Education}

Many participants discussed various impacts of COVID on themselves or their family members in the educational domain as well. In particular, many participants spoke of formal educational programs going online.  For example, a 65 year old woman from Washington state wrote: ``My son is in college and he has completely switched to online only.  He was involved in clinicals and all of them were cancelled and converted to online.''

Similarly, a woman from Florida who did not state her age wrote: ``My great grandsons are doing online classes from their school. This has been very helpful in keeping them safe.''

Only one participant, a 35 year old man from Washington, noted that schools in his area had closed but that schooling had not shifted to an online format: ``My son's school is closed down but they send packets to work on weekly, not a digital replacement.''

Beyond formal schooling, participants identified an array of informal educational activities that they or their families were undertaking. A different 26 year old man from Florida wrote that members of his family ``have started to learn new things such as new languages by using tools like Duolingo more often''.  Similarly, a 70 year old man from Texas wrote: ``I recently bought an harmonica and I availed myself of free 30 day lessons online. Pretty cool. Rather in person, but that's not free.''

Another participant's husband was using the COVID pandemic to contemplate new life options that they could undertake. The 56 year old woman from Ohio mentioned in the earlier sections noted that ``during this time my husband has been thinking of taking some classes to perhaps make a career change.'' While the connection was not made explicit by that participant, the connection between the work and education impacts on her husband's life opens the possibility that the disruption in his life potentially opened up his willingness to explore new directions.

Beyond the educational activities themselves, the shift to online education had changed some participants' perception of education more broadly. The 30 year old man from Washington state mentioned earlier wrote:
``I have been studying through a part time basis course at a local university.I am currently studying online at home using my laptop.I am able to join a zoom class hosted by one of my lecturers.Together with others in the group,computing system has become our new norm,The class is very organised and our lecturer is able to guide us through course work in an easy and efficient manner.This kind of study platform has proved to be a game changer in the education sector and it is my hope that the education stakeholders will take a leaf or two from these experiences.''

Similarly, a 33 year old man from Maryland wrote: ``The COVID19 pandemic has changed the way I view education. I have spent some time on a udemy course and am interested in more online courses. I believe that is the future of education.''

Participants had a range of emotional responses to the educational changes brought about by COVID.  For example, the 26 year old man from Florida who was annoyed by working from home was annoyed by online learning as well, writing that he ``had to finish school online. i suck at online classes [so] school was twice as hard. its annoying''.

A 33 year old woman from Florida identified challenges that online educational programs faced with certain kinds of content.  She wrote: ``I am ... a full-time student in the veterinary technology program so my classes all had to move online, including lab classes.  The lecture classes were fine, but the lab classes were much more difficult because we weren't able to go out into the field and learn on the animals but had to learn about the theory only.'' She also noted shortcomings in the readiness of teachers to teach online: ``My teenage son also had to take online classes which did not go well.  His teachers were completely unprepared and I find that most of them did not adjust to the online learning as his workload significantly dropped.''

And a  69 year old man from Florida was impressed by how his wife had engaged in self-directed education using online resources:
``My wife has done something quite extraordinary. She is retired and didn't have a lot to do. So she somehow found the information on the web on how to re-do a kitchen. Including repainting all of the doors on the kitchen cabinets and redoing all of the counter tops with this sand looking covering. There is a lot of work and it has given her something interesting to do. She gets up the first thing in the morning and gets to work. This has been an education for her.''

As with the work domain, there was ambivalence in the study population about whether the shift to more frequent computer use in education was primarily problematic-but-necessary, or if it was beneficial. In addition, a subset of participants found the changes enabled new possibilities in their lives, in particular in the domain of informal learning.



\subsubsection{Home Life}

Multiple participants identified an increase in particular kinds of home usage as a result of the COVID lockdown. Some noted a change in the nature of their activities, while others just noticed a change in the quantity/frequency of existing behaviors.

For example, the 35 year old man from Washington state wrote: ``It has increased the amount of time I spend on my computer due to video gaming and such that I do more under the lockdown/social distancing restrictions.''  A 31 year old woman from Minnesota pointed to online shopping as an activity that had increased: ``We have been avoiding going out as often an have therefore been doing a lot of online shopping.This has helped us stay safe indoors and also reduce the time we spend doing shopping.''

A 33 year old man from Maryland pointed explicitly to an increase in quantity of activity rather than the adoption of new activities, writing: ``It has not changed the way we use digital media. The only difference is the amount of time spent, since there is additional free time on our hands due to the lockdown orders.''

Similarly, the 31 year old man from New York wrote: ``I live by myself so I haven't changed how I use my computer in my home life. I might listen to music more or stream more shows but I do that in the same way as I did before, just more frequently.''

Beyond entertainment and shopping, multiple participants also noted that they used computational systems to update themselves regarding COVID-related news.  For example, a 60 year old man from Pennsylvania wrote: ``I think I am probably consuming much more digital media now than before the pandemic began as I read a lot of news on the pandemic.'' Similarly, a 41 year old man from Rhode Island wrote: ``I have been using social media more to stay in touch with friends and family. I have been sharing news stories with friends and family as well.''

Many participants identified home confinement/quarantine/lockdown as a key factor in increased home computer use. A 47 year old woman from West Virginia wrote: ``We use the computer more for our home life. Since we are confined at home, we mostly have no other alternative''. Similarly, a 40 year old woman from New York wrote: ``i use to [go to the] gym and park , but now a days we cant go any where''.

Many people mentioned specific platforms that they used for their home media consumption.  For example, a 30 year old man from Illinois wrote: ``Because I work from home, I am able to use my laptop a lot more to increase my digital media consumption. I watch Netflix, Hulu Plus, and Disney Plus from my laptop. I also watch a lot more YouTube videos as well as watching my favorite Twitch streamers. I used to do all of this before, but consumption has gone way up because of my ability to work from home due to COVID-19.'' The ``Activities, Devices, and Platforms'' section below, in particular Figure 1, provides additional detail on the specific platforms that participants mentioned.

Some participants also identified changes in which devices were being used.  For example, the 56 year old woman from Ohio mentioned earlier offered that ``now my husband is on his laptop instead of streaming content on our TV.''

The 27 year old person from New York who identified as both a woman and a man found that the changes in their life had led to problematic health effects: ``I have been using the internet and computer more because I have been staying indoors. It has been detrimental to my eye health. However, it is hard to find things to do when you are stuck at home.''

Participants identified a range of emotional factors in their computer use for home life.  For example, the 30 year old man from Washington state discussed his positive feelings toward various forms of computer entertainment, writing: ``I have been watching more movies and Tv programs now more than never before.I have even mastered the time when i should watch certain programs since most of them are very interesting and captivating Again,Video games have become part and parcel of my home life and i guess there is no single day that passes on without the excitement of a video game thrill in my own house.I also stream some exercises programs directly from you-tube.This was not my norm before COVID 19.''

A 60 year old woman from Kentucky pointed to boredom as a reason for her increased usage wrote: ``I read online more often than I used to because I am home more than i used to be. I do things inside more than i used to and when it gets very hot outside i stay inside more often and read things online also . I do it for entertainment or from boredom or for exercise or learning about something i am interested in.''

Unlike in the work or learning domains, no participants spoke unfavorably about their increase in computer use in the home life domain, perhaps due to its voluntary nature.  The entertainment and news value provided by technology was also identified as central for many participants.



\subsubsection{Social Life}

Many participants noted a significant change in how they use technology as part of their social lives. For example, the 31 year old man from New York wrote: ``I now do all of my socializing over Google Hangouts and Zoom. So instead of seeing friends in person we have a set time to hangout online. I also talk to my parents over Zoom instead of on the phone, which is a big change. I never used the computer to talk to friends or family before, it was always over the phone or text.''

Similarly, many participants mentioned that they frequently checked in on their loved ones via video chat. A 30 year old man from South Carolina wrote: ``we spend far more time doing video calls and conferences as opposed to meeting in person like we used to before. I perform welfare calls to my immediate family every morning and [my boyfriend] does the same.''

As with home life above, more time at home was a key factor in these changes, as noted by the 36 year old woman from California: ``I use social media a lot more. I go on Facebook and Instagram more since I have more time at home. I also practice social distancing, so I keep in touch with friends this way and family.''

Videoconferencing via platforms such as Zoom became a daily part of many people's social lives, such as for the 30 year old man from Illinois: ``I am on my computer a lot more which means I am on Facebook and available through email a lot more than I was before COVID-19. Also, my wife and I haven't been able to see our friends/family that we don't live with, so we use our laptops to connect to them via Zoom. We have daily video chats with friends. This has increased considerably since COVID-19. I think I only used Zoom one time before, and now have used it almost daily.''

Nevertheless, other participants pushed back against the move toward videoconferencing. For example, a 51 year old man from Michigan wrote, of his family: ``No one has jumped on the video conferencing bandwagon.  Social media, email, texting are adequate.'' 

Similarly, the 70 year old man from Texas mentioned above wrote:``For social life my computer usuage has not changed. Alsways emailed but don't use a webcam. I have called more though. Thought of buying a webcam but decided not too.'' The same participant pushed back against social media as well. ``I never used social media before so I am not going to start now. The old saying was go see them in person and if you can't then call them and if you can't send an email. Today you can add send a text but for heavens sake stay away from social media.''

Similarly, the 56 year old woman from Ohio wrote: ``I am not using social media anymore than usual, so our use hasn't changed. Besides, with all of the mis information out there on social media, as well as the complaining and in general bad vibe, I find I don't want to get on social media.

Online chat was also relevant for some people, including a woman from Florida who did not list her age: ``I find my social life is online right now. I connect and chat with people in chat rooms and that gives me someone to talk to and connect with.''

Still other participants used video games to keep in touch with their loved ones.  A 33 year old woman from Florida wrote: ``With everything being cancelled from the pandemic, my son \& I have been using video games and social media to keep in contact with friends that we would normally go and see to maintain social distancing but keep in contact with others.''

Similarly, a 42 year old man from Oregon wrote: ``My children are online gaming more to be able to connect with their friends. Especially on days that they can't communicate with them any other way.''

One participant, a 34 year old man from Florida, increased his participation in online dating: ``I have signed for some new dating services and have tried the online dates that are around now due to social distancing.''

Numerous participants had specific stories of social interactions conducted online.  For example, a 34 year old man from New York was using FaceTime to be able to meet his sister's newborn baby: ``All my family members have been using WhatsApp and FaceTime more frequently to interact with each other. My sister recently gave birth to a child so she has been contacting us through FaceTime for advice and to allow us to see her baby.'' He also pointed to the use of videoconferencing across national boundaries: ``My mother has been using WeChat to interact with people abroad and speak with people using her native language.'' 

Various people adopted new ways of socializing, or adapted old ways to new media. A 62 year old woman from Colorado wrote: ``Different types of gathering new encounters is a whole lot harder.  You can invite people over but the chairs have to be six feet apart.  Circles are the best way to see each other.   Wine night via zoom has been another inclusion.''

A 28 year old man from Kentucky described the effect on himself and two other family members, writing: ``Yeah, I know for me and my sister we have to do long-distance with our significant others now. So almost every day we are talking on Facebook messenger video chat and my mom is doing weekly Mary Kay Spa Evenings with a group of friends. These aren't things we did nearly as often before.''

Videochat sometimes enabled the rekindling of old relationships.  A 26 year old man from Maryland wrote: ``I have attended two Zoom meetings during the pandemic, one to celebrate my mother's birthday and one setup by my aunt and uncle that were used to help my father reconnect with some of his cousins that he has not seen in a long time''. 

Sometimes relationships crossed media boundaries. A 36 year old woman from New York wrote: ``[My husband and I] have both been playing more video games and using Discord to voice chat with our friends and play games together. One of our friends who doesn't play games has also joined us in chat to have some social interaction together.''

The 69 year old man from Florida pointed to the contentious side of social media: ``My wife is a registered independent and I an a very conservative. She began watching the daily Corona task force briefings and then began to see the follow up stories on the web. Totally biased and off the mark. She has been posting a lot more on Facebook exposing this hypocrisy. I have just watched from afar, occasionally adding in a jab her and there to the governors who have been messing things up royally because of their incompetence.''

The 47 year old woman from West Virginia saw online contact as the only viable choice for keeping in touch: ``We all socialize the more online these days. We dont have any other means to keep in touch with our close family and friends.''

Various participants' responses expressed affective opinions of both the impact of COVID on their social life and about the ensuing changes in their computer usage. 

Regarding the unwelcome social impacts of COVID, the 30 year old man from Washington wrote: ``I have been adversely affected when it comes to social life.With the lock down in place,i am not able to meet friends and family members as i would have wanted.I am therefore left with the option of using the digital media let alone computer programs to communicate with my friends and family.I have the whatsapp and Zoom software's in both my PC and phone and this therefore means that i am able to connect to friends and family through chatting and video conferencing.This enables me to have a feeling of a one on one kind of a meeting.''

Similarly, a man from California who did not list his age wrote: ``The COVID-19 has negatively affected my social life. To address the challenge, I use my computer to videocall friends and other family members.''

With regard to the changes in technology usage as a result of COVID, the 27 year old person from New York who identified as both a woman and a man felt that online communication was an inadequate replacement for in-person interactions: ``I have been talking to people online versus in-person for a few months now. I think that it is not the best substitute for social interaction and hope the situation improves soon.''

Nevertheless, even though computational platforms may not have been ideal for all participants, they were greatly valued by some.  For example, the woman from Florida who did not list her age, described the critical role of computing in maintaining connection with the world: ``I find I am on the computer a lot more. Researching things, playing games, chatting, editing photos, etc. It has been a lifeline to the outside world.''

The idea that technology provided a social lifeline was implicit in many participants' comments.  In addition, the role of computing as a provider of new social opportunities was another key theme that emerged in the social domain.

\subsection{Activities, Devices, and Platforms}

To address RQ2, we counted the number of times each activity, device, and platform was mentioned, the results of which are presented here.  

Participants mentioned many different computer-based activities that they engaged in as a result of COVID (see Figure 1, left). Online learning was the most commonly mentioned activity (mentioned by 34 of the 62 participants), followed by videoconferencing/videochat (29), working from home (27), and using social media (22). (We note that  several of these activities relate closely to the specific domains that we asked participants to address in particular questions in the survey. We also note that these activities do not represent mutually exclusive categories; for example, both online learning and working from home often involve videoconferencing.) Other activities included gaming (12), email (10), entertainment (10), information seeking (9), streaming video (9), consuming news (9), staying in touch (7), informal learning (6), online shopping (6), watching movies/TV (without explicitly mentioning streaming or related terms) (5), texting (5), working online (without explicitly mentioning working from home) (5), browsing the web (4), listening to music (3), online banking (2), online chat (2), and voice calls/chat (2). Numerous other activities were mentioned by only one participant (e.g., online dating, birthday party, wine night, Mary Kay Spa Evenings).

\begin{figure*}[t]
\centering
  \includegraphics[width=1\textwidth]{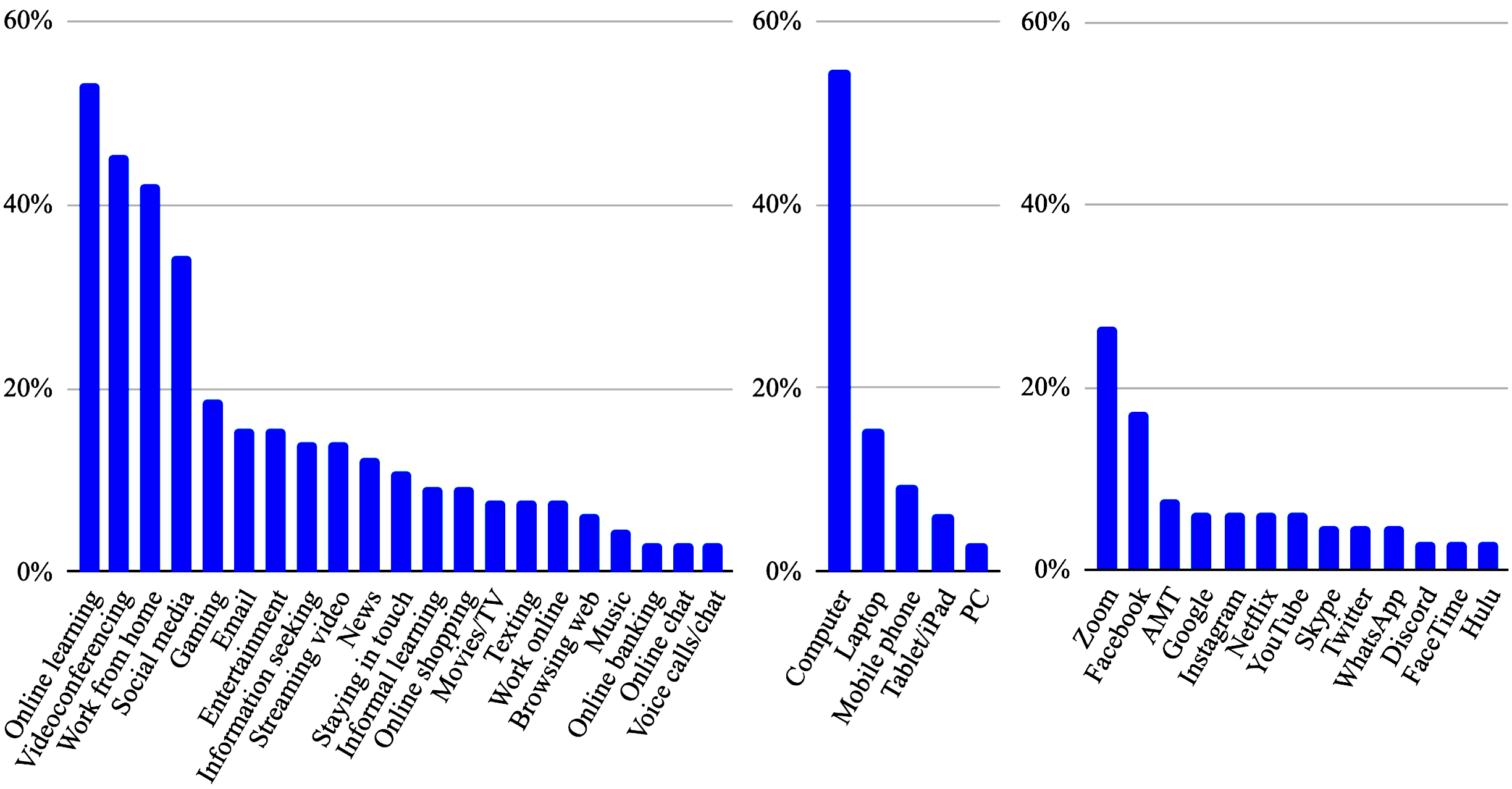}
  \caption{Charts of activities, devices, and platforms mentioned by two or more participants.  Each chart shows the percentage of participants who mentioned a given item.}
  \label{fig:allcharts}
\end{figure*}

Participants mentioned a variety of types of devices on which these activities unfolded (see Figure 1, center).  Of the devices mentioned by two or more participants (and therefore included in Figure 1), computers (without specifying laptop vs. personal computer (PC), etc.) were the most commonly mentioned devices (mentioned by 35 participants), followed by laptop computers (10), mobile phones (6), tablets/iPads (4), and PCs (2). Additional devices mentioned by only one participant included a keyboard, a mouse, a TV, and a webcam.

Participants also mentioned numerous specific platforms with which they engaged (see Figure 1, right). Of the platforms explicitly named by more than one participant, Zoom was the most common (mentioned by 17 participants), followed by Facebook (11, including one specific mention of Facebook Messenger video chat), AMT (5\footnote{We expect that AMT was mentioned as frequently as it was because the survey itself was conducted on AMT.}), Google (4, including specific mentions of Hangouts, Meet, and search), Instagram (4), Netflix (4), YouTube (4), Skype (3), Twitter (3), WhatsApp (3), Discord (2), FaceTime (2), and Hulu/Hulu Plus (2).  Each of the following was mentioned exactly once as well: Disney Plus, DuoLingo, Jabber, Microsoft Teams, Snapchat, Spotify, Telegram, Twitch, Udemy, and WeChat.  Zoom's prominence, along with the presence of several other videoconferencing platforms (e.g., FaceTime, Google Hangouts) is consistent with the earlier finding that videoconferencing was the most common activity mentioned.  Similarly, mentions of Facebook, Twitter, and Instagram are consistent with the importance of social media in the time of COVID. 

We would also like to note that the counts listed above represent how many times each item was mentioned by participants; however, in relation to RQ1, only a subset of these activities, devices, and platforms reflected a change in usage.  While some instances were new to participants, in many instances, people were continuing or amplifying pre-existing behaviors, or returning to usage patterns from previous points in their lives, or using existing technology in new ways.

\section{Discussion}
In this section we draw together several common themes that emerged across the various participants' responses.

\subsection{Changes in Computing Use}
As discussed earlier, across the various domains we asked about (work, education, home life, and social life), most participants identified increased computer usage in at least one category, and a significant subset noted increased usage across all four.  While people often noted no change in one or more areas that the survey asked about, and one person noted a reduction in one domain (as a result of an increase in another), the large majority appears to have experienced increased usage.

This increased usage arose from two main changes: the adoption of new behavioral patterns, and increased frequency of existing behaviors. In terms of new behaviors, participants spoke of themselves or their family starting to work from home, participate in online education, play games, and have videochat-based parties over the internet. In terms of increases in existing behaviors, other participants who already worked from home, played games, or used Facebook found themselves doing so more frequently.

The changes identified by participants reflect previous findings by other researchers, such as those regarding the role of videoconferencing/chatting \cite{lottridge}, the idea that online education may be a ``new normal'' \cite{triyason}, the importance of gaming \cite{amin_online_2020}, and implications for both children \cite{goldschmidt_covid-19_2020} and elders \cite{Morrow}. Given the changes participants identified in educational processes, there will likely be a growing need for new ways to create engaging experiences via remote instruction, e.g., \cite{LiAndJu}. And given the potential long-term health effects of COVID, there may also be changes in computing use as a result of future needs for rehabilitation \cite{Urban}, analogous to previous work in HCI relating to stroke patients \cite{Kelliher}.

In sum, COVID has led to both more and different forms of computing usage.

\subsection{Activities and Technologies}
Participants used numerous different platforms and devices for an even larger number of activities.  Many platforms focused on videoconferencing, and in particular Zoom. This focus was encapsulated by one participant's comments about the ``Zoom revolution''.  The Zoom revolution is also reflected in that company's stock price (ZM), which rose from US\$68.72 per share on January 2, 2020 to US\$384.48 on September 10, 2020, a more than five-fold increase. Various other videoconferencing platforms were mentioned as well, including Skype, FaceTime, and Facebook Messenger video chat. 

Beyond videoconferencing, two other substantial sets of platforms included social media (Facebook, Instagram, Twitter, etc.) and entertainment/information media (Netflix, YouTube, Hulu, etc.).  Each of these sets of platforms was mirrored by an array of related activities that participants mentioned, such as social media, online chat, streaming video, and movies/TV. 

The devices that people used were less differentiated than the activities or platforms that they mentioned. Most people simply mentioned a computer, rather than providing more detail on the nature of that computer. No one mentioned that they were on a Mac vs. Windows vs. Unix computer, for example. An iPad was mentioned once, but no other brand name devices were mentioned at all. This near-complete absence of hardware brand names, when juxtaposed with the abundance of different brand name software platforms, demonstrates the greater salience of those software platforms in people's lives.

\subsection{Necessity}
Many participants identified that the changes in their computer usage were necessary to continue previous aspects of their way of life. COVID had brought about unwelcome disruptions in many people's lives. Participants noted that various parts of their lives had been ``adversely affected'' and ``negatively affected'' by COVID, and that it had led to ``boredom'' and being ``stuck at home''. These life disruptions then led to changes in computer use.

This necessity was perhaps most pronounced in the work and education domains.  In the work domain, one participant said he ``had to change to working from home'' and another that ``COVID-19 have made it necessary for my wife to extremely use the computing technology for her various work tasks.'' In education, one participant mentioned that ``my classes all had to move online'', and another that he ``had to finish school online''.  However, the sense of necessity was also present in the home and social domains: one participant mentioned that she used the computer more because ``we mostly have no other alternative'', and another spoke of his children using a computational system because ``they can't communicate with [their friends] any other way''.  COVID caused many people to be ``stuck'' at home; these findings point to the importance of research that is focused on understanding in-home technology use and innovating in that domain, e.g., \cite{Beneteau, Luria}. 

Changes in computer use that were made due to necessity were often unwelcome.  Participants found it ``annoying and a big inconvenience'', ``much more difficult'', and ``not the best substitute for social interaction and hope the situation improves soon.'' Nevertheless, when many aspects of work, education, and social contact moved online, many people felt that they had to to engage with computational systems as a result of these shifts.


%


\subsection{Technology as Lifeline}
Social relationships were a powerful force across many participants' experiences during the pandemic. Because of the critical role of relationships in people's lives, technology was sometimes seen as a ``lifeline'' to friends and family.  Participants frequently checked in on friends and family, seeking peace of mind that their loved ones were safe, which points to the value of systems to understand the experiences that result from such COVID-driven interactions through technology, e.g., \cite{Priego}. They found themselves needing to confront challenges of long-distance relationships---to maintain existing relationships---and of dating in the new COVID world---to develop new relationships. Another participant spoke of online chat through chat rooms as a critical point of connection, ``that gives me someone to talk to and connect with''. And some participants demonstrated that people are willing to try new things (e.g., gaming, videoconferencing) to enable socialization, which resonates with investigations of the role of Pokemon Go in socialization in a world shaped by COVID \cite{LAATO2020101458}. In line with the previous discussion of necessity, many participants appear to see computing as the only option for keeping in touch with friends and family.

Similarly, various participants seemed to feel technology was a lifeline through its use for entertainment and news, ways of maintaining connections with the outside world. For example, one participant noted that he ``read a lot of news on the pandemic''. At least one participant's interest in news intersected with the social connection: ``I have been sharing news stories with friends and family as well.''

Remaining connected with family, friends, and the outside world was a substantial motivation for many participants, and technology was a crucial enabler of those connections.


\subsection{New Opportunities}
While some shifts in computing use in the time of COVID came about out of necessity, and while others were out of a strong urge for connection to loved ones and society at large, there were also more than a few instances where COVID-caused computing use opened new doors for participants.  For example, one participant's father was able to ``reconnect with some of his cousins that he has not seen in a long time''; this kind of reconnection has been explored in the literature previously, e.g., \cite{Ibarra}.  Another participant's wife undertook a significant Do-It-Yourself project, using online videos to teach herself how to complete a number of home repair projects. A third participant's mother used WeChat to connect and talk to people in her native language. Another participant took classes on Udemy, and now saw it as the ``future of education''.  And someone else took up the harmonica.

COVID also impacted people's life trajectories, such as the participant whose husband was considering a career change.  While COVID was disruptive, for this participant that disruption provided the impetus to consider making a significant personal shift and begin laying the groundwork for this career change.

Participants described these types of experiences positively, saying that it was a ``great change that [he] hope[s] is permanent'',  and that ``[his wife] kind of likes the whole idea''.

Juxtaposed with participants' negative experiences with technology usage that had been necessitated by the pandemic, the more positive affect in these cases point toward an overall sense of ambivalence toward the role of technological systems in their lives. Ambivalence toward technology is not uncommon, e.g., \cite{seo, Bala}; participants' ambivalence in this study is consistent with previous research.   While COVID itself has been devastating to many families, and COVID-caused changes in computing were frequently unwelcome for participants in this study, these instances where the change was positive rather than problematic point to the complexity of this domain.







%

\section{Limitations and Future Work}
This research has sought to lay groundwork about how COVID has affected many people's computer usage. Nevertheless, there are an array of limitations of the current study, and many useful directions for future work that are suggested by the findings presented here.  

First, this study represents only a single snapshot in time; it would be useful to conduct the same study again in the future, to study how this same population's ways of engaging  evolve over time.  Second, the study only involved US residents over the age of 18; it would be important to conduct similar studies with different participant populations (residents of other countries, children, etc.), to study the effects of COVID on computing use across the world and in different cultural and/or personal contexts.  Third, while this study sought to achieve a compromise between breadth and depth, it would be useful to study each of the four domains in greater detail as well; these additional studies could be done via interview-based studies or questionnaires to larger numbers of participants, seeking further information on themes that emerged here, such as the effects of technology on specific relationships in the time of COVID, or the affective factors in COVID-related computing use. Finally, various themes that cut across the four domains could provide interesting studies as well, such as the substitutability of email, voice, and video, or the affordances and constraints of various videoconferencing platforms.

\section{Conclusions}
COVID has changed people's lives around the world. It will continue to do so in many regions for months or years to come. Beyond COVID, other pandemics may become more frequent as well \cite{weforum2}. 

As people's lives change, their computing use often changes.  This study has sought to understand the changes in computing usage that COVID caused for one particular population. We sought to explore several domains of participants' lives, to develop a broad sense for the range of effects that unfolded.  Participants identified an array of changes to their own and their families' computer use, in terms of both the quantity and the nature of the activities undertaken. Through analyses of participants' comments, we identified instances where technology was a necessity and a lifeline, and also a source of new opportunities.

Understanding how people initially adapted to this particular pandemic may be valuable in interpreting future behavior in this domain. As subsequent pandemics or other disruptions bring about similar changes, people's computer usage will continue to evolve.  Understanding the ways changes in computing have unfolded as a result of COVID could help the field of human-computer interaction more effectively contribute to well-being and other desirable social outcomes as future disruptions affect people's lives.



\begin{acks}
The authors thank the Donald Bren School of Information and Computer Sciences at the University of California, Irvine, for their support of this research.
\end{acks}

\bibliographystyle{ACM-Reference-Format}
\bibliography{sample-base}

\end{document}